\documentclass[12pt]{iopart}

\usepackage{graphicx}
\usepackage[breaklinks=true,colorlinks=true,linkcolor=blue,urlcolor=blue,citecolor=blue]{hyperref}
\usepackage{color, graphics}
\usepackage{amssymb}

\usepackage{verbatim}   
\usepackage{subfigure}  
\usepackage{acronym,afterpage}
\usepackage{mathrsfs}
\usepackage{multirow}
\usepackage{cancel}
\usepackage{url}
\usepackage{slashed}
\usepackage{feynmp-auto}  

\usepackage{cite}

\newcommand{\eqb}{\begin{equation}}
\newcommand{\eqe}{\end{equation}}
\newcommand{\eqab}{\begin{eqnarray}}
\newcommand{\eqae}{\end{eqnarray}}
\newcommand{\nonr}{\nonumber}

\newcommand{\cL}{{\cal{L}}}

\newcommand{\cO}{{\cal{O}}}

\begin{document}

\title{FIMP Dark Matter Freeze-in Gauge Mediation and Hidden Sector}

\author{Kuo-Hsing Tsao$^{1,2,3}$}
\address{$^1$Braviant Holdings, Chicago, IL, 60606}
\address{$^2$Department of Physics,  University of Illinois at Chicago, Chicago, IL, 60607}
\address{$^3$Numina Group, Woodridge, IL, 60517}
\ead{tsaokuohsing@gmail.com}

\begin{abstract}
We explore the dark matter freeze-in mechanism within the gauge mediation framework, which involves a hidden Feebly Interacting Massive Particle (FIMP) coupling feebly with the messenger fields while the messengers are still in the thermal bath. The FIMP is the fermionic component of the pseudo-moduli in a generic metastable supersymmetry (SUSY) breaking model and resides in the hidden sector. The relic abundance and the mass of the FIMP are determined by the SUSY breaking scale and the feeble coupling. The gravitino, which is the canonical dark matter candidate in the gauge mediation framework, contributes to the dark matter relic abundance along with the freeze-in of the FIMP. The hidden sector thus becomes two-component with both the FIMP and gravitino lodging in the SUSY breaking hidden sector. We point out that the ratio between the FIMP and the gravitino is determined by how SUSY breaking is communicated to the messengers. In particular when the FIMP dominates the hidden sector, the gravitino becomes the minor contributor in the hidden sector. Meanwhile, the neutralino is assumed to be both the Weakly Interacting Massive Particle (WIMP) dark matter candidate in the freeze-out mechanism and the Lightest Observable SUSY Particle (LOSP). We further find out the neutralino has the sub-leading contribution to the current dark matter relic density in the parameter space of our freeze-in gauge mediation (FIGM) model. Our result links the SUSY breaking scale in the gauge mediation framework with the FIMP freeze-in production rate leading to a natural and predicting scenario for the studies of the dark matter in the hidden sector.
\end{abstract}

\vspace{2pc}
\submitto{\JPG}


\section{Introduction}

SUSY is widely considered to be the most compelling theory beyond the Standard Model (SM) and provides a good candidate for the WIMP dark matter, such as the neutralinos through the conventional freeze-out mechanism. However, the degeneracies of superpartners of the SM particles are not present in our visible world. Therefore, SUSY must be broken at certain energy scale $M_{\cancel{SUSY}}$. Gauge mediation is one of the most promising ways of transmitting SUSY breaking from the hidden sector to the observable world. Moreover, the null of positive results from the dark matter search suggests that dark matter might reside in the hidden sectors or non-WIMP dark matter for examples \cite{Feng:2008ya, Feng:2008mu, Ibarra:2008kn, McKeen:2009rm, Foot:2014uba}. Especially in the direct detection experiments, the scattering cross section is proportional to the strength of the dark matter interaction. None of the positive results \cite{Akerib:2016vxi,Akerib:2017kat,Cui:2017nnn,Aprile:2017iyp} leads the particle physics community to consider an alternative dark matter production mechanism, $``$freeze-in$"$, which could become favorable if dark matter direct detection result was inconsistent with the WIMP frozen-out parameter space. \cite{Hall:2009bx, Hall:2010jx, Blennow:2013jba, Cheung:2010gj, Cheung:2010gk, Cheung:2011nn, Khlopov:2012lua, Dev:2013yza, Heikinheimo:2017ofk, Bernal:2017kxu, Doroshkevich1984}.

In contrast to the conventional freeze-out process, the freeze-in production of dark matter happens while the initial relic density of dark matter is negligible and the dark matter couples to the hot bath particles which are still in thermal equilibrium at the early universe. In the freeze-in framework, the current relic density of the FIMP is determined when the temperature drops below the hot bath particle freeze-out temperature. In this paper, we explore a new scenario: the freeze-in gauge mediation (FIGM) model in which SUSY is broken by metastable vacua. The FIMP is the fermionic component of the metastable SUSY breaking pseudo-moduli which couples to the hot bath gauge mediation messengers feebly. The relic abundance of the FIMP is set by the SUSY breaking scale and the feebly coupling. In the FIGM scenario, the FIMP lodges in the hidden sector with the canonical dark matter candidate in the gauge mediation framework, the gravitino. We formularize the relic density ratio between the FIMP and the gravitino in the hidden sector and link the mass scale of the FIMP with the freeze-in production rate.

In the gauge mediation models, the gauge messenger generates the gaugino spectrum via the loop correction. Thereby the mass scale of the neutralino WIMP is linked with the messenger scale $M_{mess}$ and the SUSY breaking effect $F$ for gauge mediation models as shown in Eq.(\ref{mchi}). (See \cite{Giudice:1998bp} for more details). 
\eqb
\label{mchi}
m_{\tilde{\chi}^0} \approx \frac{g^2}{16 \pi^2} \frac{F}{M_{mess}} \sim \cO({\rm TeV})
\eqe
For $g \sim \cO(1)$, the SUSY breaking scale is bounded $\frac{F}{M_{mess} } \gtrsim $ 100 TeV from the current LHC collider limit on the WIMP mass \cite{Sirunyan:2017lae, Sirunyan:2017qaj, ATLAS-CONF-2017-039, atlassusy} and this lower bound will be further applied to constraint the parameter space of the FIMP. In addition, the gravitino is the exceptional lightest supersymmetric particle (LSP) and the dark matter candidate with R-parity conserved, and the gravitino abundance is generated by either the thermal scattering or the hot bath particles decay. The gravitino production rate is related to the reheat temperature $T_R$, the LOSP mass spectrum and the messenger scale \cite{Moroi:1993mb,Dimopoulos:1996gy,Giudice:1998bp, Choi:1999xm, Hamaguchi:2009hy,Cheung:2011nn,Fukushima:2012ra,Fukushima:2013vxa,Dalianis:2013pya}. Since the valid thermal leptogenesis requires the high reheat temperature \cite{Choi:1999xm,Fukushima:2013vxa}, $T_R > M_{mess} >M_{MSSM}$ is assumed to be higher than the Minimal Supersymmetric Standard Model (MSSM) mass scale without lack of generosity. Under this reheat temperature condition, the gravitino production from the messengers is the dominant source \cite{Choi:1999xm, Hamaguchi:2009hy, Fukushima:2012ra,Fukushima:2013vxa,Dalianis:2013pya} and the potential gravitino overproduction or the universe overclosure issue in the early universe is resolved by some mechanisms, e.g.\cite{Hamaguchi:2009hy, Fukushima:2012ra, Feng:2008zza, Dalianis:2013pya, Murayama:2007ge, Craig:2008vs, Addazi:2016mtn,Co:2017pyf}. We further show that the gravitino relic density will also be suppressed by the FIMP feeble coupling to the messengers in the hidden sector.

The FIGM links a generic metastable SUSY breaking model with the FIMP freeze-in production which provides a predictive scenario for the SUSY breaking scale from the cosmological dark matter relic density observations. The FIMP dark matter $\tilde{\chi}^h$ in the hidden sector feebly interacts with the hot bath messenger fields in the early universe and then freezes-in via the thermal scattering of hot bath particles. Furthermore, the mass of FIMP is generated by the loop correction of messengers, which is analogous to gaugino mass in the visible sector. The FIMP is the fermionic component of the SUSY breaking pseudo-moduli and receives mass after getting lifted a the quantum level \cite{Shih:2009he}.
\eqb
\label{FIMPmass}
m_{\tilde{\chi}^h} \approx \frac{\lambda_{hidden}^2}{16 \pi^2} \frac{F}{M_{mess}}
\eqe
We show that the FIMP is the LSP from Eq. (\ref{FIMPmass}) when the current dark matter density is dominated by the FIMP. On the other hand, the traditional WIMP candidates, the neutralino, now becomes the LOSP and only has the sub-leading contribution to the relic density. As mentioned earlier, the gravitino is the typical LSP in gauge mediation framework. Therefore the hidden sector becomes two-component and we estimate the ratio of the relic density between the gravitino and FIMP. This ratio is related to how SUSY breaking is commuted to the messenger sector and the feeble couplings of FIMP and we parametrize the condition in which FIMP is the majority of the hidden sector.

This paper is structured as follows. In section \ref{sec2}, we introduce and summarize the FIGM mechanism and outline the relic density of the FIMP in the parameter space of the FIGM. In section \ref{sec3}, we show the relic density ratio between the gravitino and the FIMP in the two-component hidden space. Section \ref{sec4} compares the relic density between the neutralino freeze-out and the FIMP freeze-in productions. We also show that under certain parameter space the neutralino becomes the sub-leading term in the current relic density when the FIMP has the dominant contribution. We conclude our results and outlook in section \ref{sec5}.


\section{FIMP Freeze-in Gauge Mediation}
\label{sec2}

First, we demonstrate the FIGM mechanism in which the FIMP is the fermion component of pseudo-moduli in the metastable SUSY breaking vacua. The FIMP is massless at tree level and then becomes massive from the loop contribution of gauge messengers\cite{Intriligator:2006dd,Intriligator:2007py,Haba:2007rj}. The FIMP resides in a hidden sector and feebly couples to the gauge messenger sector. This tree-level massless fermion should be distinguished from the \textit{Goldstino}, the spin-1/2  \textit{Goldstone Fermion} for the spontaneous SUSY breaking. If the Goldstino is eaten by gravity, it becomes a spin-3/2 gravitino by the Super-Higgs Mechanism \cite{Volkov:1973ix}.
The FIMP is generated by the decay of gauge mediation messengers and thermal scattering of other hot bath particles. The relic density of the FIMP is highly linked with the SUSY breaking scale and the reheating temperature of the Universe after inflation. Here, for simplicity, we assume that the gravitino relic density contribution is derived mainly from the hot bath particles and the gravitino relic density can be diluted by some mechanisms e.g. late entropy production, low reheat temperature, low SUSY-breaking scale, sequestering gravitino, etc. \cite{Choi:1999xm, Hamaguchi:2009hy, Fukushima:2012ra, Feng:2008zza, Dalianis:2013pya, Craig:2008vs}. Therefore the gravitino relic density can be the valid fraction of the dark sector. We will discuss this two-component dark sector in $\sc{Section}$ 3. Since the FIGM model is a generic metastable SUSY breaking model \cite{Intriligator:2006dd}, \textit{the dynamical evolution of the pseudo-module} \cite{Hamaguchi:2009hy, Fukushima:2012ra} naturally coincides with the metastable SUSY breaking model presented here.

We assume the FIMP superfield $X$ is the spurion-like superfield with SUSY breaking scale $F_X$.  The standard gauge mediation SUSY breaking messengers ($\Phi$, $\tilde{\Phi}$) commute with both the MSSM sector and the hidden sector. There is no direct mediation between the MSSM and the hidden sectors shown in Eq. (\ref{WFIGM}) and the FIGM interactions in Eq. (\ref{LFIGM}): 
\eqab
\label{WFIGM} 
W = \lambda_x X \tilde{\Phi} \Phi + \lambda_{z_i} Z_i \tilde{\Phi} \Phi + W_{\rm OR} (X,~Z_i)   \\
\label{LFIGM}
\cL_{FIGM} =  \lambda_x \phi \bar{\psi} \tilde{\chi}^h + \lambda_x \tilde{\phi} \bar{\tilde{\psi}} \tilde{\chi}^h + h.c. 
\eqae
where Eq. (\ref{WFIGM}) is a generic O'Raifeartaigh SUSY breaking model. $Z_i$ are the hidden SUSY breaking chiral superfields (the SM singlets) and will not directly participate in the FIGM mechanism. The FIMP is generated from the thermal scattering of hot bath particles within the messenger portal and the decay of messengers in Eq. (\ref{LFIGM}). We will show the latter makes the sub-leading contribution in the current relic density.

Note that one of the $Z_i $ is the \textit{canonical moduli} which breaks SUSY spontaneously and its fermion component is a massless \textit{Goldistino} except the appearance of gravity mentioned above. Here, we assume SUSY is broken by meta-stable vacua \cite{Intriligator:2006dd}. The $\lambda_{z_i} Z_i \tilde{\Phi} \Phi $ term provides the gravitino production channels from the decay and scattering of messengers similar to FIMP. For simplicity, we assume $i=1$ from now on. $\Phi$ and $\tilde{\Phi}$ are the gauge mediation messengers which have the SM-charged scalar and fermion components $(\phi, \psi), (\tilde{\phi}, \tilde{\psi})$. The dark matter is the fermionic component noted as $\tilde{\chi}^h$ which is stabilized by R-parity. The R-parity assignment is shown in TABLE. \ref{tab:Rparity} where both scalar and fermion masses of $X$ are generated by the loop correction of the messengers :
\eqab
m_{\tilde{\chi}^h} \sim \frac{\lambda_x^2}{16 \pi^2} \left( \frac{\lambda_x F_X}{M_\phi} \right)
\label{FIMP-mass} \\
M^2_{S} \sim \frac{\lambda_x^2}{16 \pi^2} \left(\frac{\lambda_x F_X}{M_\phi} \right)^2 
\eqae 
and we also assume $\sqrt{\lambda_x F_X} \ll M_\phi$ for simplicity.

\begin{table}[htbp!]
\begin{center}
\caption{\label{tab:Rparity}
The $R$ parity assignment of the chiral superfields in FIGM model. }
\begin{tabular}{ccccc}
\hline
\textrm{} & \textrm{$X$} & \textrm{$Z$}  & \textrm{$\Phi$}  & \textrm{$\tilde{\Phi} $}\\
\hline
Scalar   &$S$        &$G$    & $\phi$      & $\tilde{\phi} $  \\
R-parity  &$+$     &$+$          & $-$          & $-$   \\
Fermion   &$\tilde{\chi}^h$  &$\tilde{G}$       & $\psi$      & $\tilde{\psi} $\\
R-parity                          &$-$      &$-$        & $+$           & $+$        \\
\hline
\end{tabular}
\end{center}
\end{table}

\begin{figure}[!ht]
\centering
\begin{subfigure}[FIGM mode $\sharp1$:  Gauge messenger decays into the LSP.]{
\begin{fmffile}{LOSPdecay1}
\begin{fmfgraph*}(160,100)
\fmfleft{i1}
\fmfright{o1,o2} 
\fmf{scalar,label=$\phi$}{i1,v1}
\fmf{quark,label=$\tilde{\chi}^h$,label.side=right}{v1,o1}
\fmf{quark,label=$\bar{\psi}$}{o2,v1}
\end{fmfgraph*}
\end{fmffile}
\label{fig:LOSPdecay1} }
\end{subfigure}
\vspace{5mm}
\begin{subfigure}[FIGM mode $\sharp2$: Bath particles freeze-out and the LSP freeze-in.]{
\begin{fmffile}{LOSPscattering}
\begin{fmfgraph*}(160,100)
\fmfleft{i1,i2}
\fmfright{o1,o2} 
\fmf{quark,label=$\tilde{\lambda}$,label.side=left}{i1,v1}
\fmf{quark,label=$\bar{\psi}$,label.side=left}{v1,i2}
\fmf{scalar,label=$\phi$,label.side=right}{v1,v2}
\fmf{quark,label=$\tilde{\chi}^h$,label.side=left}{v2,o1}
\fmf{quark,label=$\bar{\psi}$,label.side=left}{o2,v2}
\end{fmfgraph*}
\end{fmffile}
\label{fig:LOSPscattering} }%
\end{subfigure}
\label{fig:FIGM}                            
\caption[FIGM mechanism leading terms]{The leading order of the FIGM mechanism has two modes:  (a) The decay of the scalar component of the gauge mediation messenger into the hidden sector (b) FIMP $\tilde{\chi}^h$ freezing into the hidden sector by thermal scatterings of the hot bath particles, e.g. the gaugino $\tilde{\lambda}$ and the fermionic component of messengers.}
\end{figure}
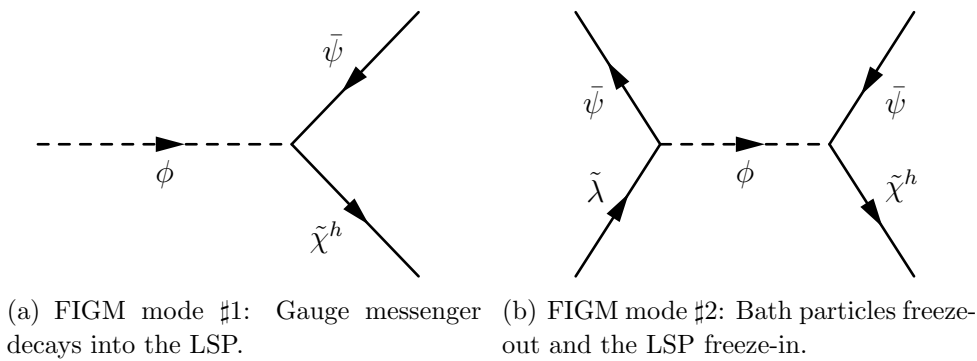

Within the freeze-in model, we assume $\tilde{\chi}^h$ has negligible initial thermal abundance in the early universe and that feebly couples to thermal hot bath when both the messengers and the MSSM particles are still in the thermal equilibrium.\footnote{The scalar partner of FIMP, $S$, is generated by freeze-in first and then decay to di-jet and di-photon via messenger loop. The effect on the relic density from the late decay of $S$ is negligible.}When the reheat temperature $T_R$ is still higher than the mass of hot bath particles ($T_R > M_\phi >M_{MSSM}$ ), the yield of $\tilde{\chi}^h$ is produced by the decay of the messenger field and the thermal scattering of the MSSM particles illustrated in Figure \ref{fig:LOSPdecay1} and \ref{fig:LOSPscattering}:
\eqab
\label{y1to2}
Y^{\tilde{\chi}^h}_{1 \rightarrow 2} &\approx&  \frac{135 \times g_\phi}{8 \pi^3 \times 1.66 \times g^S_{*} \times \sqrt{g^{\rho}_{*}}} \times \frac{M_{Pl} \Gamma_{\phi \rightarrow \psi \tilde{\chi}^h}}{M^2_\phi}\\
\label{y2to2}
Y^{\tilde{\chi}^h}_{2 \rightarrow 2} &\approx&  \frac{135 \times g^2 \lambda_x^2}{512 \pi^5 \times 1.66 \times g^S_{*} \times \sqrt{g^{\rho}_{*}}} \times \frac{M_{Pl}}{\Gamma^{\rm total}_{\phi}}\
\eqae
where $g$ is the gauge coupling of MSSM particle to messengers,  $g_\phi$ is degree of freedom of $\phi$, $g^S_{*}$ and $g^{\rho}_{*}$ are the degree of freedom in the hot bath at freeze-in temperature $T_{FI} > M_{\phi}$ for the entropy $S$, the energy density $\rho$, $M_{Pl} \approx 10^{18} \rm GeV$ and the decay widths are:
\eqab
\Gamma_{\phi \rightarrow \psi \tilde{\chi}^h} &\approx& \frac{1}{8\pi} \lambda_x^2 M_\phi \left(1 - \frac{M_\psi}{M_\phi} \right)^2 = \frac{\lambda_x^4}{8 \pi} \frac{F_X^2}{M_\phi^3}\\
\Gamma^{\rm total}_\phi &\approx&  \frac{1}{8\pi} \left(g^2 + \lambda_x^2 \right)  M_\phi  \approx \frac{g^2 }{8\pi} M_\phi,~g \gg \lambda_x 
\eqae
where $M_\phi^2~=~M_\psi^2 \pm \lambda_x F_X$ ($M_\phi^2 \gg \lambda_x F_X$).

We show that when $\lambda_x \ll \cO(1)$ and $\lambda_x F_X \ll M_\phi^2$, the ratio of FIMP density between the decay and the scattering of the hot bath particles is:
\eqb
\frac{Y^{\tilde{\chi}^h}_{1 \rightarrow 2}}{Y^{\tilde{\chi}^h}_{2 \rightarrow 2}} \approx \left( \lambda_x^2 \frac{F_X}{M_\phi^2} \right)^2 \ll 1
\eqe
which implies that the scattering process is the dominant channel for FIMP freeze-in relic abundance. 

\begin{figure}
\centering
\includegraphics[scale=0.45]{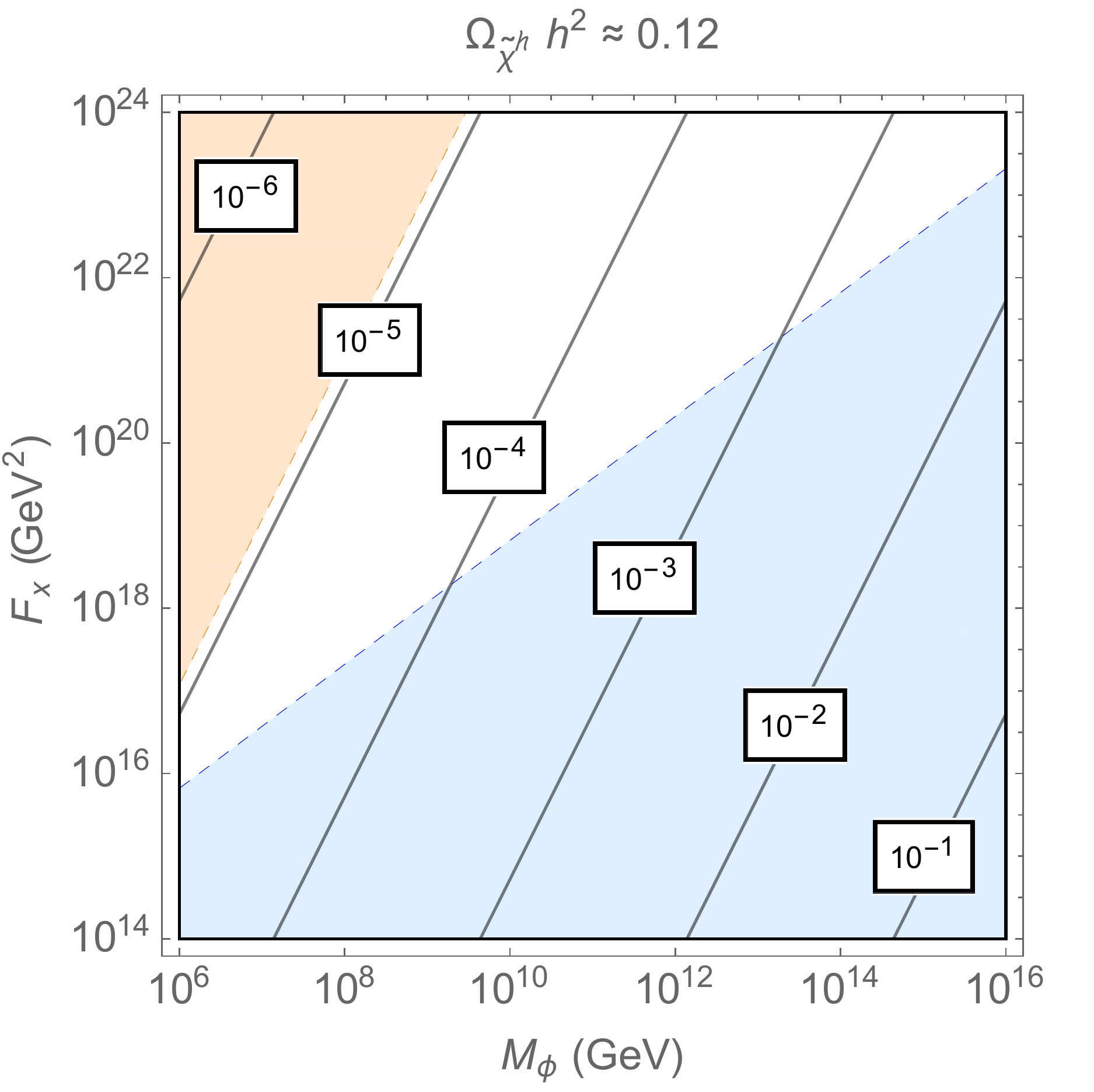}
\caption[Feeble coupling for FIMP relic abundance]{The contour plot of feeble coupling $\lambda_x$. The {\sc framed box} shows the value of $\lambda_x$ in order for the dark matter relic abundance $\Omega_{\tilde{\chi}^h} h^2 \approx 0.12$ with respect to different SUSY breaking effects $F_x$ and messenger masses. The lower-right {\sc blue} region shows the exclusive region $\lambda_x F_X / M_{\phi} \leq 100~\rm TeV$ from current LHC limits and the upper-left {\sc organge} region is due to $\lambda_x F_X > M_\phi^2$ assumption. }
\label{fig:fimp-relic}
\end{figure}

On the other hand, if $T_R < M_\phi$ then the messengers are not in the thermal bath which leads the thermal scattering in Figure \ref{fig:LOSPscattering} to become an off-shell process \cite{Elahi:2014fsa} and thus the number density becomes:
\eqb
Y^{\tilde{\chi}^h}_{2 \rightarrow 2,~\rm off-shell} \approx \frac{90 \times g^2 \lambda_x^2}{128 \pi^7 \times 1.66  \times g^S_{*} \times \sqrt{g^{\rho}_{*}}} \frac{M_{Pl} T_R^3}{M_\phi^4}
\eqe
Here we focus on the former case, i.e. $T_R > M_\phi$ and the relic abundance of $\tilde{\chi}^h$ from the scattering of messengers:
\eqb
\Omega_{\tilde{\chi}^h} = \frac{S_0}{\rho_c} m_{\tilde{\chi}^h} Y_{{\tilde{\chi}^h}} 
\label{FIMP-relic}
\eqe
where $S_0 \approx \rm 2.8 \times 10^3~cm^{-3}$ is the current entropy and critical density $\rho_c \approx 10^{-5} h^2 \rm ~GeVcm^{-3}$. Since $m_{\tilde{\chi}^h}$ is generated by the messenger loop correction Eq. (\ref{FIMP-mass}), we can further write Eq. (\ref{FIMP-relic}) in terms of SUSY breaking scale $F_X$ and messenger mass $M_\phi$:
\eqab
\Omega_{\tilde{\chi}^h} h^2 &\approx& \frac{2.8 \times 10^8}{\rm GeV} m_{\tilde{\chi}^h} Y_{{\tilde{\chi}^h}} \nonr \\
 &\approx& 0.12 \times \left( \frac{\lambda_x}{10^{-5}} \right)^5 \times \left(\frac{F_{X}}{10^{16}~\rm GeV^2} \right) \times \left(\frac{10^6 ~\rm GeV}{M_\phi} \right)^2 \nonr \\
\label{FIMP-relic-full}
\eqae
where $M_{Pl} \approx 10^{18}~\rm GeV$, $g_\phi =1$ and $g^S_{*}$ and $g^{\rho}_{*}$ are about $\cO(10^2)$. The MSSM gauginos also receive masses from messenger loop correction and the current LHC lower bound limits on gaugino masses give us $\lambda_x F_X / M_{\phi} \gtrsim 100~\rm TeV$. We present the feeble coupling with respect to different messengers and the SUSY breaking scale in Eq. (\ref{WFIGM}) in order to have FIMP relic abundance  $\Omega_{\tilde{\chi}^h} h^2  = \Omega_{DM} h^2 \approx 0.12$ shown in Figure \ref{fig:fimp-relic}. As you can see, $\lambda_x$ is about $\cO(10^{-5})$ to $\cO(10^{-4})$ in the inclusive $F_X$ and $M_\phi$ parameter space from the current LHC $\lambda_x F_X / M_{\phi} \leq 100~\rm TeV$ limit and the $\lambda_x F_X < M_\phi^2$ assumption.

To summarize FIGM mechanism, we begin with a generic metastable SUSY breaking model in the gauge mediation framework:
\begin{itemize}
\item R-parity \cite{Chamseddine:1982jx, Barbieri:1982eh} makes pseudo-moduli FIMP $\tilde{\chi}^h$ stable.
\item The tree-level mass of pseudo-moduli FIMP is forbidden by R-symmetry. \cite{Shih:2009he}
\item $\tilde{\chi}^h$ obtains mass at one-loop correction from integrating out the massive messengers.
\item Messengers are in thermal equilibrium with the hot bath particles ($T_R > M_\phi$), and $\tilde{\chi}^h$ is produced from the hot bath particles scattering and resides in the SUSY breaking sector with feebly coupling to the messengers.
\item The initial negligible relic abundance of the FIMP $\tilde{\chi}^h$ increases due to the feebly interaction with the hot bath particles and then the relic density of FIMP freezes in when the hot bath particles freeze out.
\end{itemize}


\section{Two-component Hidden Sector}
\label{sec3}

The gravitino (the typical LSP of gauge mediation models) resides in the hidden sector along with the FIMP. Both the gravitino and the FIMP have the same production modes shown in Figure \ref{fig:LOSPdecay1} and \ref{fig:LOSPscattering}: the thermal scattering and decay of the messengers. Hence the hidden sector becomes two-component.\footnote{In the conditon of $T_R > M_\phi$, the production of gravitino from other hot bath particles is highly suppressed \cite{Choi:1999xm}.} Due to the mass generation method difference and how \textit{canonical moduli} and \textit{pseudo-moduli} feels the SUSY breaking effect, we further show that the two-component hidden sector phenomenon is linked with the SUSY breaking scales and the feeble coupling of the FIMP. 

First, we define k as the Goldstino portion of SUSY breaking effect:  
\eqb
k \equiv {F_Z \over F} = \frac{F_Z}{ \sqrt{F_X^2 + F_Z^2}} \approx {F_Z \over F_X},~ F_X \gg F_Z
\eqe
where $F_X$, $F_Z$, and $F$ are the FIMP pseudo-module, the gravitino module and the total SUSY breaking effect respectively. $\lambda_x$ and $\lambda_z$ are the hidden (the FIMP) and the visible sector (the gravitino) couplings to the messengers respectively. Hence, the gravitino mass leads to:
\eqb
m_{3/2} = \frac{F_Z}{k \sqrt{3} M_{pl}}
\eqe
and the gravitino relic density is dominated mainly by the thermal scattering of the messengers. The MSSM sector has a negligible contribution when reheat temperature is larger than $M_{mess}$. We also ignore the model dependent superWIMP scenario \cite{Choi:1999xm, Feng:2003uy, Cheung:2011nn, Dalianis:2013pya}.

The interaction between the gravitino and the messengers
\eqb
\cL = \lambda_z k \phi \bar{\psi} \tilde{G} + \lambda_z k \tilde{\phi} \bar{\tilde \psi} \tilde{G} + h.c.
\eqe
which leads to the  gravitino number density ratio between the decay and the scattering from messengers as shown below:
\eqb
\frac{Y^{\tilde{G}}_{1 \rightarrow 2}}{Y^{\tilde{G}}_{2 \rightarrow 2}} \approx \left( \lambda_z^2 k^2 \frac{F_X}{M_\phi^2} \right)^2 \ll 1.
\eqe
Similar to the FIMP, the gravitino abundance is also dominated by the hot bath particles scattering rather than the decay.

Therefore, the relic density of gravitino is:
\eqb
\Omega_{\tilde{G}}  h^2 \approx 0.1 \times \left( \frac{\lambda_z}{10^{-2}} \right)^4 \times \left( \frac{F_{Z}/M_\phi}{10^2 ~\rm GeV} \right) \times \left( \frac{k}{10^{-3}} \right)
\label{Gravtino-relic}
\eqe

Now we realize the relic density of the hidden sector is composed of the sum of gravitino relic density in Eq. (\ref{Gravtino-relic}) and FIMP in Eq. (\ref{FIMP-relic}). The two-component hidden sector can be presented by the ratio as shown below:
\eqb
\frac{\Omega_{\tilde{G}} h^2 }{\Omega_{\tilde{\chi}^h} h^2 } \approx  \left( \frac{10^{-5}}{\lambda_x} \right)^4  \times \left( \frac{\lambda_z}{10^{-2}}\right)^2  \times \left( \frac{k}{10^{-3}} \right)^2 \times \left( \frac{M_\phi} {10^6~{\rm GeV}}\right)
\label{ratio-hs}
\eqe
We can see that the ratio between the gravitino and the FIMP is controlled by the freeze-in couplings and $k$ for a given gauge messenger mass scale.

\begin{figure}[htbp!]
\centering
\includegraphics[scale=0.4]{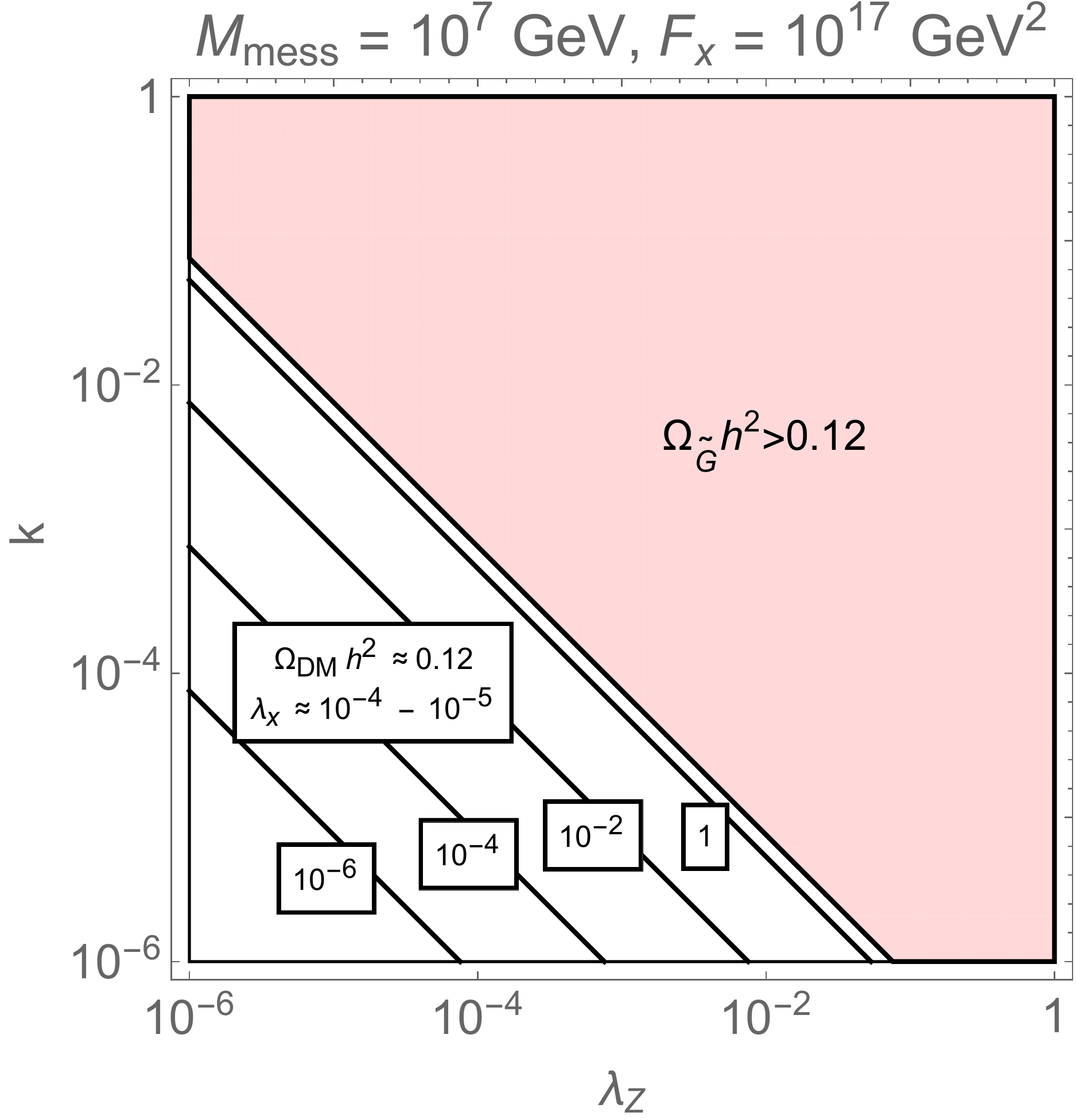}
\caption{The asymmetric hidden sector behavior of the relic density in the plot of the coupling of gravitino to messengers and $k$ at one branching point $M_{\phi}=10^7~{\rm GeV}$ and $F_x=10^{17}~{\rm GeV^2}$. The number in the {\sc framed box} is the ratio of $\frac{\Omega_{\tilde{G}} h^2 }{\Omega_{\tilde{\chi}^h} h^2 }$. The {\sc Red} region exclusively shows gravitino overproduction parameter space when $k$ and $\lambda_z$ larger than $\cO(10^{-1})$. The {\sc White} region shows the ratio of gravtino to FIMP $\Omega_{\tilde{G}}$/$\Omega_{\tilde{\chi}^h}$ while the combination of gravitino and FIMP relic density $\Omega_{DM}h^2 =\Omega_{\tilde{G}}h^2~+~\Omega_{\tilde{\chi}^h} h^2 \approx 0.12$ and $\lambda_x \approx 10^{-5}$. }.
\label{fig:ratio-hs}
\end{figure}

Here we illustrate the two-component hidden sector relic density $\Omega_{DM}=\Omega_{\tilde{\chi}^h}+\Omega_{\tilde{G}}$ by considering a branching point $T_R \gg M_{
\phi}=10^7~{\rm GeV}$ and $F_x=10^{17}~{\rm GeV^2}$ in Figure \ref{fig:ratio-hs}. In particular, the gravitino overproduces when $k$ and $\lambda_z$ are about $\cO(1)$.  As you can see, when $k$ or $\lambda_z$ becomes infinitesimal the FIMP dominates the hidden sector, which means the SUSY breaking effect is mainly dominated by the FIMP sector $F \approx F_{X}$. The masses of the FIMP and gravitino are about $\cO(\rm  KeV)$ and $\cO(\rm 10^2~MeV)$ respectively at this branching point of parameter space. The mass hierarchy between FIMP and gravitino in the parameter space of FIGM shown in Figure \ref{fig:fimp-relic}:
\eqb
\frac{m_{\tilde{G}}}{m_{\tilde{\chi}^h} } \approx 10^5 \left( \frac{M_\phi} {10^6~{\rm GeV}}\right) \times  \left( \frac{10^{-5}}{\lambda_x} \right)^3
\eqe

We conclude the relic density ratio of the two-component hidden sector with Eq. (\ref{Gravtino-relic}) and (\ref{ratio-hs}) in which the FIMP dominates the hidden sector majorly in the FIGM parameter space of $F_X$ and $M_\phi$ from Figure \ref{fig:fimp-relic} and SUSY breaking effect at $F_X \gg F_Z$ or parametrically $k < 10^{-2}$ and $\lambda_z < 10^{-2}$.

\section{LOSP Neutrino Freeze-Out Versus FIMP Freeze-In}
\label{sec4}

When the FIMP and the gravitino freeze-in from the decay of gauge messengers or the scattering between the neutralino and chargino, the neutralino ($\tilde{\chi}^0$) freeze-out process also might contribute to the relic density of dark matter when the neutralino is assumed to be the lightest observable SUSY particle (LOSP) in the visible sector. Here we estimate the portion of the neutralino in the dark matter relic density by considering the tree level annihilation of the neutralinos. In contrast to the freeze-in process, the neutralino freezes-out with the co-moving volume number density:
\eqb
Y_{\tilde{\chi}^0} \approx \frac{1}{\left < \sigma_{ann} v\right> M_{pl} m_{\tilde{\chi}^0}}~,~\left< \sigma_{ann} v \right>  \approx  \frac{g_{a}^4}{16 \pi^2 m_{{\tilde{\chi}^0}}^2}
\label{LOSP-relic}
\eqe
where $g_a$ is the neutralino gauge coupling in the annihilation process. We can further rewrite Eq.(\ref{LOSP-relic}) with neutralino mass generated by the loop correction of the messengers so that the relic density of the neutralino leads to Eq. (\ref{LOSP-relic-full})
\eqab
 m_{\tilde{\chi}^0}&\approx&\frac{g_n^2}{16 \pi^2} \frac{\lambda_x F_x}{M_\phi} \approx 1~{\rm TeV} \times \left( \frac{g_n}{\cO(1)} \right)^2  \times \left( \frac{\lambda_x F_x/M_{\phi}}{10^5~{\rm GeV}} \right)
 \label{LOSP-mass} \nonr \\
 \Omega_{\tilde{\chi}^0} h^2 &\approx& 0.017  \times \left(\frac{\lambda_x F_x/M_{\phi}}{10^5~{\rm GeV}} \right)^2 \times \left(\frac{g_n}{g_a} \right)^4 
\label{LOSP-relic-full}
\eqae
where $g_n$ is the gauge coupling of the neutralino. As $g_n \sim g_a$, the neutralino relic density is the sub-leading term compared to Eq. (\ref{FIMP-relic-full}), however, the ratio between the FIMP and the neutralino is determined by the SUSY breaking scale, the messenger mass scale and the FIMP feeble coupling $\lambda_x$ as shown in Eg. (\ref{fimp-vs-lops}).
\eqb
\frac{\Omega_{\tilde{\chi}^0} h^2}{\Omega _{\tilde{\chi}^h} h^2} \approx \left( \frac{10^{-5}}{\lambda_x} \right)^4 \times \left( \frac{\lambda_x F_x/M_{\phi}}{10^5~{\rm GeV}} \right) \times \left( \frac{M_\phi} {10^6~{\rm GeV}}\right)
\label{fimp-vs-lops}
\eqe

\begin{figure}[htbp!]
\centering
\includegraphics[scale=0.4]{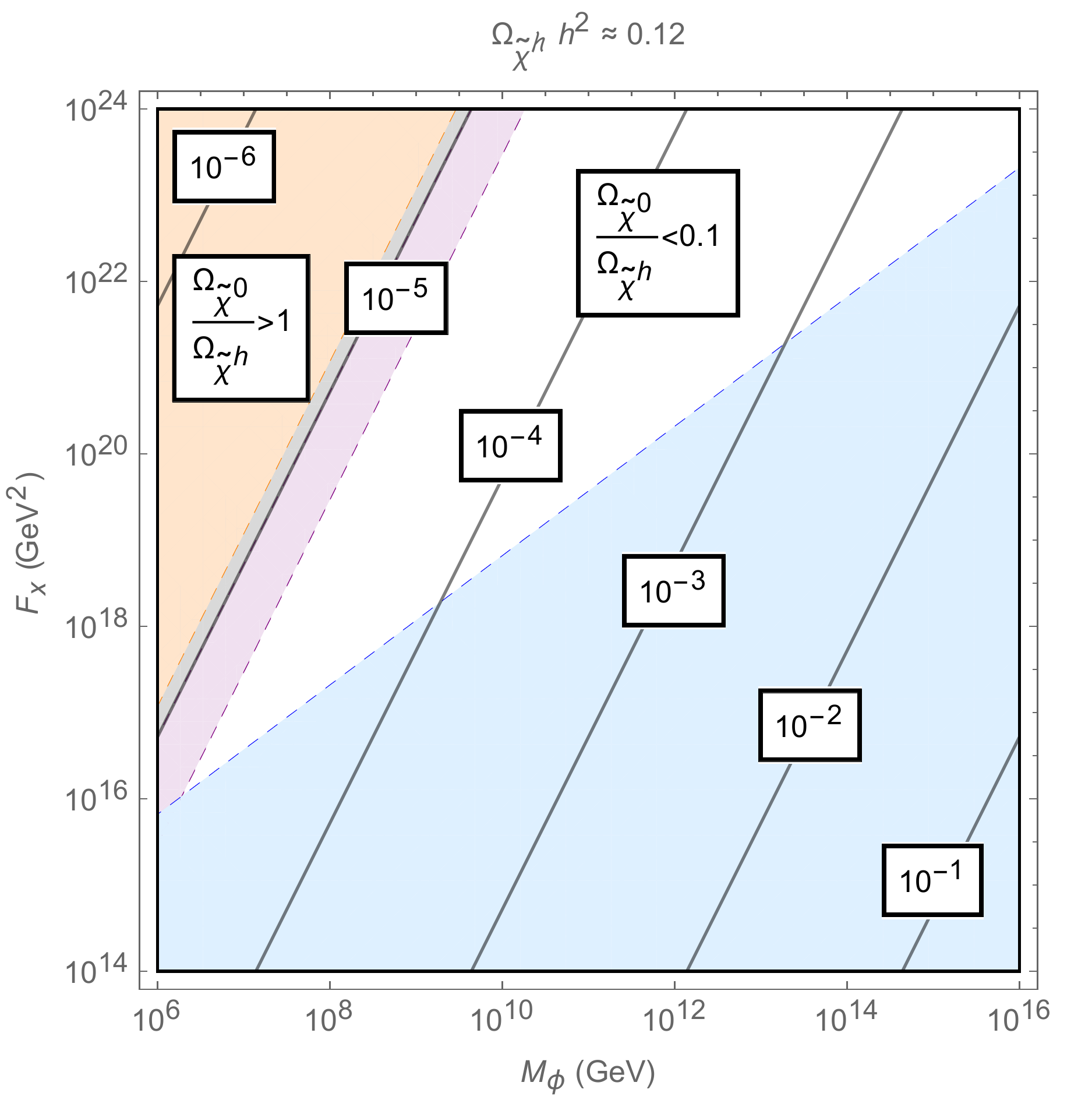}
\caption{We show the ratio between the FIMP and the neutralino (LOSP) at different SUSY breaking scales and messenger masses where $\lambda_x$ is chosen to satisfy $\Omega_{\tilde{\chi}^h} h^2~\approx~0.12$. The {\sc Grey} region is excluded when LOSP has more relic density than currently observed dark matter relic density. The {\sc Pink} region is where LOSP becomes the sub-leading term compared to FIMP: $0 < \frac{\Omega_{\tilde{\chi}^0} h^2}{\Omega _{\tilde{\chi}^h} h^2}  <1$. The {\sc White} region shows where FIMP dominates the hidden sector and neutralino relic density is negligible. The lower-right {\sc blue} region shows the exclusive region $F_X / M_{\phi} \leq 100~\rm TeV$ from current LHC limits.}
\label{fig:fimp-losp-full}
\end{figure}

We present the ratio of the neutralino relic density to the FIMP relic density and $\lambda_x$ is chosen to make $\Omega_{\tilde{\chi}^h} h^2 ~\approx~0.12$ in Figure \ref{fig:fimp-losp-full}. As it shows that the neutralino has sub-leading or negligible relic density when $\lambda_x \approx 10^{-5}$ and neutralino mass is about few TeV.

\section{Summary and Outlook}
\label{sec5} 
 
In this paper, we explore the FIMP dark matter freeze-in gauge mediation (FIGM) model. The FIMP is a metastable SUSY breaking pseudo-moduli fermion feebly coupling to the gauge messengers and receives mass from the loop correlation of the messengers. When the messengers are still in the hot bath at the early universe and in the condition of $T_R > M_\phi$, the FIMP is created mainly through the scattering of the hot bath particles via the messenger mediator shown in Figure \ref{fig:LOSPscattering} and the FIMP relic density freezes-in when the temperature drops below the hot bath particle mass. The accessible FIGM parameter space includes the feeble coupling between the FIMP and messengers, the messenger scale, and the SUSY breaking scale based on the current LHC searching limits on the neutralino mass presented in Figure \ref{fig:fimp-relic}. We further compare the relic density between the FIMP and the gravitino which is the expected hidden sector particle in the gauge mediation framework. We find out the gravitino becomes the sub-leading or even negligible contributor in the hidden sector when the SUSY breaking effect is mainly dominated by the FIMP field which becomes a new strategy to suppress the gravitino relic density. Both the FIMP freeze-in and the neutralino freeze-out processes summarize the FIGM parameter space. We conclude that the FIMP dominates the most of hidden sector relic density when the parameters fit in Figure \ref{fig:ratio-hs} and \ref{fig:fimp-losp-full}. 

The FIGM provides a natural and alternative scenario for the future hidden sector studies especially when conventional WIMP parameter space is facing severe challenges from the current dark matter searches. Particularly when the physics community begins to suggest the dark matter direct detection searches for KeV to MeV dark matter mass range \cite{Essig:2011nj, Essig:2012yx,Lee:2015qva,Essig:2015cda,Izaguirre:2015yja,Green:2017ybv}, which coincides the FIMP mass range discussed here. The FIMP mass range is also optimal for the dark matter indirect detection scenarios like the nuclear beta decay and the electron capture processes \cite{Biermann:2006bu, deVega:2009ku,  deVega:2011xh, Moreno:2016hrs, Adhikari:2016bei}. Moreover, the FIGM links the dark matter density with SUSY breaking scale, and FIMP mass scale which leads to an alternative but highly predictable direction for the future SUSY and dark matter searches. In a forthcoming paper, the details of the connection between dynamical SUSY breaking models and FIMP relic density will be discussed.


\section{Acknowledgments}

We gratefully acknowledge C.-H. Chang, I. Dalianis, A. Ismail, W.-Y. Keung, T.-C. Kuo, S. Seto and S. Sharp for useful discussions and to J. Unwin for the early cooperation.

\section*{References}
\bibliographystyle{vancouver}
\bibliography{FIGM_JPG.bib}
\end{document}